\documentclass[12pt,final]{article}
\usepackage{amsmath,amssymb,amsfonts,amsthm,slashed,braket}
\usepackage[all]{xy}
\usepackage{mathrsfs}
\usepackage{fancyhdr}
\usepackage{showkeys}
\usepackage{hyperref}

\newcommand\nc{\newcommand}
\nc\linesep{\bigskip}
\nc\newprob[1]{\marginnote{#1}[\parskip]}
\nc\bA{\mathbb A}
\nc\bC{\mathbb C}
\nc\bD{\mathbb D}
\nc\bR{\mathbb R}
\nc\bZ{\mathbb Z}
\nc\bQ{\mathbb Q}
\nc\bP{\mathbb P}
\nc\bV{\mathbb V}
\nc\bW{\mathbb W}
\nc\bG{\mathbb G}
\nc\mf\mathfrak
\nc\mc\mathcal
\nc\mb\mathbb
\nc\brac[1]{\langle#1\rangle}
\nc\abs[1]{\lvert#1\rvert}
\nc\norm[1]{\lVert#1\rVert}
\nc\onto{\twoheadrightarrow}
\nc\into{\hookrightarrow}
\nc\lto{\longrightarrow}
\nc\action{\curvearrowright}
\nc\on\operatorname
\nc\wbar\overline
\nc\what\widehat
\nc\wtilde\widetilde
\nc\nop\DeclareMathOperator
\nc\eps{\varepsilon}
\nc\tsym{\widetilde{\text{Sym}}}
\nc\oarrow[1]{\overset{#1}\to}
\nop\Hom{Hom}
\nop\End{End}
\nop\Aut{Aut}
\nop\im{Im}
\nop\id{id}
\nop\tr{Tr}
\nop\coker{coker}
\nop\Spec{Spec}
\nop\Jac{Jac}
\nop\Ext{Ext}
\nop\Tor{Tor}
\nc\op{\text{op}}
\nop\loc{Loc}
\nop\Frac{Frac}
\nc\ann{\text{ann}}
\nop\QCoh{QCoh}
\nop\Coh{Coh}
\nop\Sym{Sym}
\nop\Hilb{Hilb}
\nop\gr{Gr}
\nop\Tot{Tot}
\nop\Fl{Fl}
\nop\tGamma{\widetilde\Gamma}
\nop\tloc{\widetilde{\text{Loc}}}
\nop\rep{Rep}
\nop\proj{Proj}
\nc\oo[1]{\overset\circ{#1}}
\nop\ospec{\oo{Spec}}
\nop\oTot{\oo{Tot}}
\nop\Bl{Bl}
\nop\Comp{Comp}
\nop\Ho{Ho}
\nop\cone{Cone}
\nop\LKE{LKE}
\nop\RKE{RKE}
\nop\pd{pd}
\nop\cd{cd}
\nop\depth{depth}
\nop\ass{Ass}
\nop\supp{supp}
\nop\codim{codim}
\nop\holim{\underset{\lto}{holim}}
\nop\dlim{\underset{\lto}{lim}}

\nop\uHom{\underline{\Hom}}
\nop\Pic{Pic}
\nop\Cl{Cl}
\nop\Div{Div}
\nop\rank{rank}
\nop\Der{Der}
\nop\dimrel{dim.rel}
\nc\sHom{\mathscr Hom}
\nc\sExt{\mathscr Ext}
\nc\dto{\dashrightarrow}
\nop\rspec{\bf Spec}
\nop\Gal{Gal}
\nop\Ind{Ind}
\nop\Frob{Frob}
\nop\Fib{Fib}
\nop\ratdim{rat\ dim}
\nop\Mod{Mod}
\nop\rat{rat}
\nop\val{val}
\nop\Rep{Rep}
\nop\colim{colim}
\nop\ind{ind}

\theoremstyle{theorem}

\theoremstyle{remark}

\newcommand{\bea}{\begin{eqnarray}}
\newcommand{\eea}{\end{eqnarray}}
\newcommand{\bee}{\begin{eqnarray*}}
\newcommand{\eee}{\end{eqnarray*}}
\newcommand{\al}{\begin{align*}}
\newcommand{\eal}{\end{align*}}
\newcommand{\be}{\begin{equation}}
\newcommand{\ee}{\end{equation}}
\newcommand{\bem}{\begin{pmatrix}}
\newcommand{\eem}{\end{pmatrix}}

\newcommand{\bb}{\mathbb} 
\renewcommand{\th}{\theta}

\def\({\left(}
\def\){\right)}
\def\nn{\nonumber}
\numberwithin{equation}{section}

\begin{document}

\centerline{\large{Counting spinning dyons in maximal supergravity:}}
\medskip
\centerline{\large{
The Hodge-elliptic genus for tori}}
\bigskip
\bigskip
\centerline{Nathan Benjamin$^1$, Shamit Kachru$^1$ and Arnav Tripathy$^2$}
\bigskip
\bigskip
\centerline{$^1$Stanford Institute for Theoretical Physics}
\centerline{Stanford University, Palo Alto, CA 94305, USA}
\medskip
\centerline{$^2$Department of Mathematics, Harvard University}
\centerline{Cambridge, MA 02138, USA}
\bigskip
\bigskip
\begin{abstract}

We consider $M$-theory compactified on $T^4 \times T^2$ and describe the count of spinning $1/8$-BPS states.  This refines the classic count of Maldacena-Moore-Strominger in the physics literature and the recent mathematical work of Bryan-Oberdieck-Pandharipande-Yin, which studied reduced Donaldson-Thomas invariants of abelian surfaces and threefolds. As in previous work on $K3 \times T^2$ compactification, we track angular momenta under both the $SU(2)_L$ and $SU(2)_R$ factors in the 5d little group, providing predictions for the relevant motivic curve counts. 
%The resulting $1/8$-BPS counting function is a refined avatar of the ratio of degree two Siegel modular forms $\chi_{12} / \chi_{10}$.
%We also note a connection of our results to instanton partition functions appearing in brane engineering of supersymmetric field theories.
\end{abstract}

%MMS, BOPY, KKP-esque moonshine

%should figure out Georg's email

%extra symmetry in the Vafa field-theory limit implicit already in MMS (through the enhanced E_{6,6}(Z) symmetry?)

%does MMS already do everything here

%what is going on in 0912.0057

%maybe we should make some comments here that we're only trying to treat the torsion one multiplicity one case or what have you due to 0908.0039 being so careful about various subtleties
\newpage
\tableofcontents

% 0. jumping 1. Lerche (+some moonshine?) 2. short fully-flavored K3? 3. \Phi_{12} in K3 \times T^4 enumerative geometry 4. CHL lift paper? 5. more expansions and singular theta lift 6. D1-D5 indefinite theta function 7. Nakajima ... (genus two, classification, more attractor/flux)

\section{Introduction}
\label{sec:sec1}

The counting of microstates contributing to BPS black hole entropy in $K3 \times T^2$ compactification of M-theory started with the work of Strominger-Vafa in \cite{SV}, and was given a more precise description in \cite{DVV} shortly thereafter. The seemingly easier, and more supersymmetric, case of a purely toroidal compactification of $M$-theory took a few more years until the treatment of \cite{MMS}. Indeed, the computation for the $\mc{N} = 4$ theory arising from $K3 \times T^2$ compactification\footnote{We use the convention that half-maximal supersymmetry in 5d is called ${\cal N}=4$, while maximal supersymmetry is ${\cal N}=8$; this is in keeping with the normal convention in 4d.} proceeded via computing elliptic genera of the family of CFTs one obtains from the $D1-D5$ system, and these analogous counts in the $\mc{N} = 8$ theory naively vanish. It hence took some ingenuity to define an appropriately-corrected elliptic genus in order to perform a nontrivial computation. 

\medskip
Recently, a Hodge-elliptic genus was proposed as an analogous quantity that would be of use in computing not just the BPS spectrum in such theories, but also its flavoring by the full $SU(2) \times SU(2)$ little group of the theory \cite{KT} (while typically, earlier approaches kept only a single $SU(2)$ quantum number).  In \cite{KT}, the $\mc{N} = 4$ theory above was treated.  Here, we offer an analogous treatment of the $\mc{N} = 8$ theory. This case of maximal supersymmetry is in some sense a nicer showcase
for the Hodge-elliptic genus, in that the unflavored counting function suffers from subtleties alluded to above: the extra supersymmetry forces the need for recurring modifications to the counting formulas and techniques used.  On the other hand, the flavored counting function from the Hodge-elliptic genus can be evaluated straightforwardly in the ${\cal N}=8$ theory, with no need for special modifications that do not arise in cases with less supersymmetry. Of course, we remind the reader that the Hodge-elliptic genus badly fails to be an index and may jump discontinuously as one moves in moduli space \cite{jumping}.  While this upper semicontinuous jumping is entirely physical in that the flavored BPS spectrum indeed varies with moduli, one may reasonably object to this added complication if one is only interested in the unflavored count. Note furthermore that in this case, 
the relevant worldsheet theory is a $\sigma$ model with target $\Hilb^n T^4$.  
Maldacena-Moore-Strominger were certainly able to solve for the full partition function of this free theory, and worked directly from this even more informative function before reducing to their supersymmetric index.  In this free theory, by studying the Hodge-elliptic genus we are essentially simply insisting on focusing attention on a less specialized limit of the easily computable full partition function.  We will leave aesthetic deliberations regarding these approaches to the judgment of the reader.

\medskip
Mathematically, as in \cite{KT}, the effect of flavoring our BPS particle count by the additional $SU(2)$ angular momentum is to refine a Donaldson-Thomas generating function to a motivic Donaldson-Thomas generating function \cite{DimofteGukov}. We hence provide an interpretation of our results here in this language in section \ref{sec:sec4}, extending conjectures of \cite{Bryan}. Mathematical readers may hence wish to only briefly skim the intervening sections for the electrifying thrill before focusing on section \ref{sec:sec4}.

\medskip
The next section recalls the BPS spectrum of this $\mc{N} = 8$ theory, largely following \cite{MMS} and \cite{SSY, Pioline, Sen}.  We present the refined counts in section \ref{sec:sec3}. Section \ref{sec:sec4} provides the mathematical interpretation in terms of motivic Donaldson-Thomas invariants.
% In section \ref{sec:sec5}, we briefly discuss relations of our formulae to instanton partition functions in Nekrasov's $\Omega$ background, which have been computed in the geometric engineering literature.

\section{The unflavored $1/8$-BPS spectrum}
\label{sec:sec2}

%Various prefactors are probably wrong in all formulas. Possibly some mention somewhere should be made of the 4d vs 5d counts and the Shih-Strominger-Yin N = 8 paper interpolating between them, but I can't find actual expressions for all these counts anywhere -- maybe you have it in your papers/talks? I am also generally confused which functions below are inverted and which are not -- various things seem to not be very much in analogy with the K3 case. The additive lift formula seems to think we are lifting a weight 2 Jacobi form, not weight -2. 

We study the D1-D5 system on $T^4 \times S^1$, and consider the computation of BPS states on the worldvolume of the resulting effective string.
In contrast to the $\mc{N} = 4$ case, here the counting function capturing the $1/4$-BPS spectrum is just $1$.  If we attempt to count Dabholkar-Harvey states \cite{DH}, we have particles in the ground state on the right and the excitations of $8$ bosonic and $8$ fermionic oscillators on the left.  These precisely cancel.  More formally, where in the ${\cal N}=4$ theory we would obtain in this way a sum of Euler characteristics of $\Hilb^n K3$, now we wish to sum \be\sum \chi(\Hilb^n T^4) q^n = 1.\ee The sum is $1$ by localization, as the manifolds occurring in each term except the zeroth term admit a free $T^4$-action, and hence have trivial Euler characteristic. In other words, the indexed count of $1/4$-BPS states only captures the vacuum. 

\medskip
In fact, the more refined counts, such as the $1/4$-BPS spectrum flavored by the $SU(2)_L$ angular momentum or the $1/8$-BPS spectrum (which in a sense already has the $SU(2)_L$ flavoring), would also be trivial if not corrected. In order to provide a nontrivial match to black hole entropy, the authors of \cite{MMS} performed a more sophisticated count of these states by weighting the counts by $F_R^2$. If one computes these corrected counts by considering the $D1-D5$ frame (using $U$-duality to suppose we have a single $5$-brane and some variable number of $1$-branes), considering the effective field theory of the $1$-branes dissolved in the $5$-brane yields the $\sigma$-model to $\Hilb^n T^4$. The relevant counts here are naively given by the $\chi_y$ genus and the elliptic genus $Z_{EG}$, respectively.  To obtain the more sophisticated counts, which we denote by the reduced $\tilde{\chi}_y$ genus and the reduced elliptic genus $\tilde{Z}_{EG}$, we need again insert an $F_R^2$ in the worldvolume CFT traces on the effective string. In fact, we only need to know these traces for the first theory, the $\sigma$-model to $T^4$.  The other CFTs of interest are simply its orbifold symmetric powers, and any trace over the full series will be given by a multiplicative lift of the answer for the first CFT \cite{DMVV}.

\medskip
The reduced $\tilde{\chi}_y$ genus is easy to compute.  Recalling the full Hodge diamond of $T^4$, with Hodge polynomial \begin{align} \text{Hodge}(T^4) &= (y^{1/2} - y^{-1/2})^2(u^{1/2}-u^{-1/2})^2 \nn \\ &= y^{-1}u^{-1} - 2y^{-1} - 2u^{-1} + y^{-1}u + 4 + yu^{-1} - 2u - 2y + yu  \nn \\ \Longrightarrow \tilde{\chi}_y(T^4) &= -(1 \cdot 0^2 - 2 \cdot 1^2 + 1 \cdot 2^2) y^{-1} + (2 \cdot 0^2 - 4 \cdot 1^2 + 2 \cdot 2^2) - (1 \cdot 0^2 - 2 \cdot 1^2 + 1 \cdot 2^2) y \nn \\ &= -2y + 4 - 2y^{-1}.\end{align} 

\medskip
%We
%may multiplicatively lift this result to find the reduced spinning $1/4$-BPS state count \begin{eqnarray} \sum (c_n^{r_L})_{5d} F_R^2 p^n y^{[r_L]} &=& \sum_n \tilde{\chi}_y(\Hilb^n T^4) p^{n-1} \nn \\ &=& \prod_{n=1}^{\infty} \frac{(1 - p^ny)^2(1 - p^ny^{-1})^2}{(1 - p^n)^4} \nn \\ &=& \( \frac{ \prod_{n=1}^{\infty} (1-p^ny^{-1})(1-p^n)(1-p^ny) }{ \prod_n (1 - p^n)^3 } \)^2 \nn \\ &=& - \( \frac{\theta_1(\sigma, z)}{\eta(\sigma)^3} \)^2 . \end{eqnarray} Here, we use the notation $(c_n^{r_L})_{5d}$ for the number of $1/4$-BPS representations with spin $r_L$ under $SU(2)_L$ in five-dimensions, which we need to weight by two factors of the right-moving fermion number to cancel fermion zero-modes.  We also use the notation \be j^{[\ell]} = j^{-2\ell} + j^{-2(\ell - 1)} + \cdots + j^{2(\ell - 1)} + j^{2\ell}\ee to track characters of $SU(2)$. 

% I think the equation above is wrong; I think the more better one is below--

We may multiplicatively lift this result to find the reduced spinning $1/4$-BPS state count
\begin{align} \frac12\sum (c_n^{r_L})_{5d} &F_R^2 p^n y^{[r_L]} = \frac12\sum_n \tilde{\chi}_y(\Hilb^n T^4) p^{n} \nn \\ 
&= \frac12\(u \partial_u\)^2\(\prod_{n=1}^{\infty} \frac{(1 - p^ny)^2(1 - p^ny^{-1})^2(1-p^n u)^2(1-p^n u^{-1})^2}{(1 - p^n)^4(1-p^n uy)(1-p^n u^{-1}y)(1-p^n uy^{-1})(1-p^n u^{-1}y^{-1})}\)\Bigg|_{u=1} \nn \\ 
&= (y^{-1} - 2 + y)p + (2y^{-2} + y -6 + y + 2y^2)p^2 \nn \\ & \phantom{aaaaaaaaa}+ (3y^{-3} + y^{-1} - 8 + y + 3y^{3})p^3 + \mc{O}(p^4). 
 \end{align}
 Here, we use the notation $(c_n^{r_L})_{5d}$ for the number of $1/4$-BPS representations with spin $r_L$ under $SU(2)_L$ in five-dimensions, which we need to weight by two factors of the right-moving fermion number to cancel fermion zero-modes.  We also use the notation \be j^{[\ell]} = j^{-2\ell} + j^{-2(\ell - 1)} + \cdots + j^{2(\ell - 1)} + j^{2\ell}\ee to track characters of $SU(2)$. 

\medskip
Note that in the above multiplicative lift procedure (unlike the $K3$ case), we have to take some care with fermionic versus bosonic modes, placing them in the numerator or denominator appropriately. %The answer we get here at end of the chain of equalities above happens to be (up to a sign) the named Jacobi form $\phi_{-2, 1}(\sigma, z)$ (see e.g. \S4.3 of \cite{DMZ}, eq (4.29)),
%% times a prefactor $p(-y+2-y^{-1})$ (PROBABLY wrong)
%where we use the standard notation when working with
%automorphic forms:
%\be p = e^{2 \pi i \sigma}, y = e^{2 \pi i z}~.\ee
%If we instead performed the count of spinning $1/4$-BPS states in 4d, the center-of-mass degrees of freedom in the 4d-5d lift would no longer contribute and then we would have simply $$\sum (c_n^{r_L})_{4d} F_R^2 p^n y^{[r_L]} = \phi_{-2, 1}(\sigma, z).$$

\medskip
We could in fact compute the reduced $1/8$-BPS spectrum directly, using the idea that $1/8$-BPS particles are dyons of two $1/4$-BPS particles \cite{JS}, thereby writing the generating function as an additive lift of the reduced spinning $1/4$-BPS spectrum count (in fact, overall as a reduced analog of the Maass-Skoruppa lift for the reduced unflavored $1/4$-BPS spectrum count, which is still just $1$). Here, in fact, the reduced elliptic genus\footnote{Normalized as in eq (2.1) of \cite{SSY}.} is, up to a sign, the named Jacobi form $\phi_{-2,1}$ (see e.g. \S4.3 of \cite{DMZ}, eq. (4.29)) \be\tilde{Z}_{EG}(T^4) = -\phi_{-2, 1} =  \sum c(n, \ell) q^n y^{\ell}\ee and its (correctly reduced) multiplicative lift as per Dijkgraaf-Moore-Verlinde-Verlinde (DMVV) \cite{DMVV} gives the reduced $1/8$-BPS state count \be\Phi_{\rm 5d} = \sum_{n \ge 1, m \ge 0, \ell} \frac{c(nm, \ell) p^n q^m y^{\ell}}{(1 - p^n q^m y^{\ell})^2}.\ee 
%Again, this is a prefactor of $p\phi_{-2, 1}(\sigma, z)^{-1}$ (STILL probably wrong) times the count in 4d. 

\medskip
In fact, an application of the generating function identity \be\frac{t}{(1 - t)^2} = \sum_{n=0}^{\infty} nt^n\ee yields that the reduced $1/8$-BPS state count may be rewritten as \begin{eqnarray} \Phi_{\rm 5d} &=& \sum_{n, m, \ell, s} s c(nm, \ell) p^{ns} q^{ms} y^{\ell s} \nn\\ &=& \sum_{n \geq 1, m \geq 0, \ell} ~\sum_{d|(n, m, \ell)} d c\(\frac{nm}{d^2}, \frac{\ell}{d}\) p^n q^m y^{\ell},\end{eqnarray} so that in this case we see the reduced multiplicative lift of the reduced elliptic genus actually coincides with this kind of additive lift expression. 

\medskip
The formula above tells us that the entropy of states with $Q_1$ D1-branes wrapping $S^1$, $Q_5$ D5-branes wrapping $T^4 \times S^1$,  $m$ units of momentum on the circle, and
$\ell$ units of $SU(2)_L$ angular momentum is given by
\be\Omega_{\rm 5d}(Q_1,Q_5,m,\ell) = \sum_{d|(n,m,\ell)} d ~c\({Q_1 Q_5 m\over d^2},{\ell\over d}\).\ee
Note that this formula holds for mutually co-prime charges.\footnote{By mutually co-prime, we mean that no single factor divides all of the charges.  The reason for the subtlety in cases with non co-prime charges is that the relevant D-brane moduli space contains multi-center components, rendering the analysis considerably more subtle.}
%  The authors of \cite{MMS} in fact conjecture an extension of this formula to general
%charges:
%\be\tilde \Omega_{\rm 5d}({n,m,\ell}) = \sum_{d} d ~N(d)~ c\({nm\over d^2},{\ell\over d}\)\ee
%where $N(d)$ is the number of divisors of 
%\be Q_1,Q_5,m, d, {mQ_1 \over d},{mQ_5 \over d},{Q_1Q_5\over d}, {mQ_1Q_5\over d^2}~.\ee

\medskip
Now, we perform a 4d/5d lift to find the 4d counting function, following \cite{SSY,Pioline,Sen}.  
The result is
\be\Phi_{\rm 4d} = \sum_{n \geq 0, m \geq 0, \ell}{c(nm,\ell)p^n q^m y^{\ell} \over (1-p^n q^m y^{\ell})^2}~,\ee
differing from the 5d result by the inclusion of the $n=0$ term.  The sum over $\ell$ should be taken to run
over only $\ell > 0$ when $n=m=0$.

\section{The flavored $1/8$-BPS spectrum}
\label{sec:sec3}

As in prior work in the $K3$ case \cite{KT, KKP}, we may refine the above counts by flavoring by the $SU(2)_R$ angular momentum. In this more supersymmetric case, this refinement carries the additional benefit that we no longer need to take some sort of reduced, sophisticated count in order to find nonvanishing BPS generating functions. Instead, all our counts proceed in complete analogy with the $K3$ case. 

\subsection{Refined counts}
\label{sec:refcount}

\medskip
We first return to the $1/4$-BPS particle spectrum, now flavoring by both $SU(2)_L$ and $SU(2)_R$. Putting everything back in the $D1-D5$ frame, we find that we are computing Hodge polynomials of the respective $\sigma$-model targets and evaluate, by the logic of \cite{DMVV}, 

\begin{align}
\sum (c_n^{r_L, r_R})_{5d} p^n y^{[r_L]} u^{[r_R]} &=  \sum \text{Hodge}(\Hilb^n T^4) p^n \nn \\ &= \prod_{n = 1}^{\infty} \frac{(1 - y^{-1}p^n)^2(1-u^{-1}p^n)^2(1-yp^n)^2(1-up^n)^2}{(1-y^{-1}u^{-1}p^n)(1-y^{-1}up^n)(1-p^n)^4(1-yu^{-1}p^n)(1-yup^n)}~.
\end{align} 
We can again write this as the prefactor \be-\frac{1}{16} \frac{u-y-y^{-1}+u^{-1}}{u_-^2 y_-^2}\ee
times the multivariate Jacobi form 
\be\varphi(\sigma,\nu,z) = \frac{ \theta_1(\sigma, z)^2 \theta_1(\sigma, \nu)^2 }{ \theta_1(\sigma, z + \nu) \theta_1(\sigma, z - \nu) \eta(\sigma)^6},\label{eq:varphi}\ee
where we define $y = e^{2\pi i z}, u = e^{2 \pi i \nu}, p = e^{2\pi i \sigma}$, and the notation \begin{eqnarray} u_- &=& \frac{u^{-1/2} - u^{1/2}}{2},\nn \\ y_- &=& \frac{y^{-1/2} - y^{1/2}}{2}. \end{eqnarray} 

\medskip
We now move to the spinning $1/8$-BPS state count. As in \cite{KT}, the five-dimensional count is given by the multiplicative lift of the Hodge-elliptic genus $Z_{HEG}$, defined as \be Z_{HEG} = \tr_{\text{right g.s.}} \((-1)^F q^{L_0 - c/24} y^{F_L} u^{F_R}\),\ee where the trace is taken over the subspace of the Ramond-Ramond Hilbert space where the right-moving part is a ground state. This definition, applied to a $\sigma$-model to $T^4$, gives a function $Z_{HEG}(T^4)$ that a priori may depend heavily on the $T^4$ in question.  And indeed, it does: there are visibly points in the moduli space of the $T^4$ where we may pick up extra chiral currents and $Z_{HEG}$ will jump (upper semi-continuously). In \cite{KT}, however, the Hodge-elliptic genus was computed at a generic point in moduli space for a torus in any dimension, as reconfirmed there by a mathematical sheaf cohomology computation that should pick out the large-volume (generic) answer. We recall the generic answer \be Z_{HEG}(T^4) = -\( 4 \frac{\theta_1(\tau, z)}{\theta_1^*(\tau, 0)} u_- \)^2.\ee Here, \be\theta_1^*(\tau, 0) = -2q^{1/8} \prod_{n=1}^{\infty} (1 - q^n)^3\ee is essentially a provocative way of writing $\eta(\tau)^3$, so that in the above we have the usual (indexed) answer $\phi_{-2, 1}(\tau, z)$ with some prescribed polynomial dependence in $u$. 

\medskip
As stated above, the five-dimensional spinning $1/8$-BPS state count is a multiplicative lift of the Hodge-elliptic genus.  At some point in moduli space, given \be Z_{HEG}(T^4) = \sum c(n, \ell, k) q^n y^{\ell} u^k,\ee we have \be {\Phi^{\rm refined}_{\rm 5d}(\sigma,\tau,\nu,z) } = \sum (c_{n, \ell, m}^{r_R})_{5d} p^n y^{\ell} q^m u^{[r_R]} = \prod_{n \ge 1, m \ge 0, \ell, k} (1 - p^n y^{\ell} q^m u^k)^{-c(nm, \ell, k)}.\ee Here $c_{n, \ell, m}$ is the count of $1/8$-BPS states with $n=Q_1 Q_5$ (the product of the numbers of D1 and D5 branes), $\ell$ giving the $SU(2)_L$ angular momentum, $k$ giving the $SU(2)_R$ angular momentum, and
$m$ counting the momentum on the circle.  By analogy with the familiar picture of ${\cal N}=4$ black holes, it may also be convenient to think of the quantum numbers other than
$SU(2)_R$ as electric and magnetic charges, via
\begin{align}
n &={1\over 2} Q_e \cdot Q_e \nn\\
m &= {1\over 2} Q_m \cdot Q_m\nn\\
\ell &= Q_e \cdot Q_m
\end{align}

\medskip
Note that the counting in $\Phi_{\rm 5d}^{\rm refined}$ is again valid for mutually co-prime charges.  
We will be content to work at this level of generality, but it is important to remember that there would be two natural extensions.
The BPS count admits further flavoring to keep track of individual $U(1)$ symmetries instead of just U-duality invariants.  And as the U-duality symmetry in five dimensions is
$E_{6,6}({\mb Z})$, there should be automorphic forms for $E_{6,6}$ which play a natural role in the theory.  See e.g. \cite{Pioline} for further discussion of this aspect.

\medskip
As usual, and as studied in detail for this theory in \cite{SSY,Pioline,Sen}, we may also recover the four-dimensional state count via the 4d/5d lift.
In terms of the multiplicative lift, this adds back the $n=0$ term that is absent in $\Phi^{\rm refined}_{\rm 5d}$, yielding
\be{\Phi^{\rm refined}_{\rm 4d}(\sigma,\tau,\nu,z) } = \sum (c_{n, \ell, m}^{r_R})_{5d} p^n y^{\ell} q^m u^{[r_R]} = \prod_{n \ge 0, m \ge 0, \ell, k} (1 - p^n y^{\ell} q^m u^k)^{-c(nm, \ell, k)}. \label{eq:thisguy}\ee
It is to be understood in taking the product that when $n=m=0$, one should restrict to $\ell < 0$.
The resulting 4d count takes the form
\be{\Phi^{\rm refined}_{\rm 4d}(\sigma,\tau,\nu,z)} = {1\over 4} {\varphi(\tau,\nu,z) \over u_{-}^2}~{{\Phi^{\rm refined}_{\rm 5d}(\sigma,\tau,\nu,z)}}\label{eq:thisotherguy}\ee
Once again, a natural extension would be to promote this to an $E_{7,7}({\mb Z})$ invariant expression to respect the U-duality of the 4d theory; in the unrefined case, such an expression was provided in
\cite{Senexceptional}.

%Can we say anything better about the U-duality invariants?
\medskip
The above invariants do reduce back to the invariants of \cite{MMS} in a suitable limit of parameters, but in a slightly sophisticated way.  If one simply unflavors the $SU(2)_R$ angular momentum by taking $u \to 1$, the counts simply vanish.  In order to obtain the nontrivial counts with the $F_R^2$ insertion, we note following the definition of the Hodge-elliptic genus that
\begin{align}
\frac12 \(u \frac{\partial}{\partial u}\)^2 Z_{HEG} \Big|_{u=1} &= \frac12 \(u \frac{\partial}{\partial u}\)^2 \tr_{\text{right g.s.}}\Big( (-1)^F q^{L_0 - c/24} y^{F_L} u^{F_R} \Big) \Big|_{u=1} \nn \\ 
&= \frac12 \tr_{\text{right g.s.}} \Big( (-1)^F q^{L_0 - c/24} y^{F_L} (F_R)^2 u^{F_R} \Big) \Big|_{u=1} \nn \\ 
&= \frac12 \tr_{\text{right g.s.}} \Big( (-1)^F q^{L_0 - c/24} y^{F_L} (F_R)^2 \Big) \nn \\ 
&= \frac12 \tr \Big( (-1)^F (F_R)^2 q^{L_0 - c/24} y^{F_L} \Big).
\end{align}
Notice that the last step, where we replace the trace over the sub-Hilbert space of states with right-moving part a ground state (all in the Ramond-Ramond sector) with the full Hilbert space, only works given sufficient supersymmetry and fermion zero-modes to make the usual index vanish, as is the case here. One can check explicitly now that our refined counts can be simplified back to the original count of Maldacena-Moore-Strominger \cite{MMS} yielding $\Phi_{\rm 5d}$, or the expression of Sen for 
$\Phi_{\rm 4d}$ \cite{Sen}, by applying $\frac{\partial^2}{\partial \nu^2}$ to the appropriate refined counting function and taking $\nu \to 0$.

Finally we note that because our refined count is not an index, and is computed at the symmetric orbifold point where $g_s=0$ in the gravity dual arising in AdS/CFT, we are not counting black hole entropy. It is possible that cancellations occur as we move away from the orbifold point, and the black hole entropy is smaller as one moves away (see e.g. \cite{heg-sugra}).

\subsection{$SL(2,\bb Z)$ invariance}

We now discuss automorphy properties of $\Phi_{\rm 4d}^{\text{refined}}$. In particular we show that $\Phi_{\rm 4d}^{\text{refined}}$ exhibits invariance under an $SL(2,\bb Z)$ similar to the one which preserves $\Phi_{\rm 4d}$,
as discussed in \cite{Sen} (where it is related to S-duality). The $SL(2,\bb Z)$ action is
\be
\Phi_{\rm 4d}^{\text{refined}}(\sigma', \tau', z', \nu') = \Phi_{\rm 4d}^{\text{refined}}(\sigma, \tau, z, \nu)
\ee
where
\begin{align}
\sigma' &= d^2 \sigma + b^2 \tau + 2bd z\nn\\
\tau' &= c^2\sigma + a^2 \tau + 2ac z\nn\\
z' &= cd\sigma+ab\tau+(ad+bc)z \nn\\
\nu' &= \nu
\end{align}
and $\begin{pmatrix} a & b \\ c & d \end{pmatrix} \in SL(2,\bb Z)$. To show this, we will prove invariance under both $S = \bem 0 & -1 \\ 1 & 0\eem$ and $T=\bem 1 & 1 \\ 0 & 1\eem$.

\medskip
The $S$ transform takes 
\begin{align}
\sigma' &= \tau \nn\\
\tau' &= \sigma \nn\\
z' &= -z
\end{align}
which exchanges $p$ and $q$, and takes $y$ to $y^{-1}$. From the definition of $\Phi^{\text{refined}}_{\rm 4d}$ in (\ref{eq:thisguy}), it is clear that there is a $p, q$ exchange symmetry. Exchanging $y$ with $y^{-1}$ is charge conjugation which is also a symmetry present.

\medskip
Now note that the $T$ transform takes
\begin{align}
\sigma' &= \sigma + \tau + 2z \nn\\
\tau' &= \tau \nn\\
z' &= \tau + z.
\label{eq:spectralflow}
\end{align}
This takes $y$ to $yq$ and $p$ to $p q y^2$. Recall that we can write $\Phi^{\text{refined}}_{\rm 5d}$ as \cite{KT}
\be
\Phi^{\text{refined}}_{\rm 5d}(\sigma, \tau, z, \nu) = \sum_{n=0}^{\infty} p^n Z_{HEG}(\text{Sym}^n(T^4))(\tau, z, \nu).
\ee
From this we see that acting on $\Phi^{\text{refined}}_{\rm 5d}$, (\ref{eq:spectralflow}) has a clear interpretation as spectral flow on the left by one unit, which we recall takes
\begin{align}
L_0 &\rightarrow L_0 + J_0 + c/6 \nn\\
J_0 &\rightarrow J_0 + c/3.
\end{align} 
Thus $\Phi^{\text{refined}}_{\rm 5d}$ is invariant under (\ref{eq:spectralflow}). Now, to show $\Phi^{\text{refined}}_{\rm 4d}$ is invariant, we just need to show 
\be
\varphi(\tau,\nu, z) = \varphi(\tau,\nu, z + \tau).
\label{eq:pigbicycle}
\ee
(see (\ref{eq:thisotherguy})).

\medskip
From the classic identity
\be
\th_1(\tau, z+\tau)q^{1/2}y = -\th_1(\tau, z)
\ee
and the definition of $\varphi(\tau, \nu, z)$ in (\ref{eq:varphi}), we see that (\ref{eq:pigbicycle}) is satisfied, proving $SL(2,\bb{Z})$ invariance of $\Phi^{\text{refined}}_{\rm 4d}$.

\section{Motivic DT invariants of abelian varieties}
\label{sec:sec4}

%I'm confused about something here -- where in the [BOPY] paper do they have the F_R^2 weighting that allows them to get a nonzero indexed count?

%At some point I should also figure out Georg's email about the multiple covers contributing by some polynomial instead of an exponential or whatever -- almost certainly it has to do with the annoyance of extra supersymmetry and our multiplicative lift formulas looking more like additive lift formulas and whatnot. 

We recall the enumerative geometry interpretation of refining by the $SU(2)_R$ angular momentum, following \cite{DimofteGukov}. The refined invariants of the prior section assume an interpretation as motivic Donaldson-Thomas invariants, refining the enumerative geometry interpretations found in \cite{Bryan}
for the indexed counts, notably in their Corollary $5$. The doubly-spinning $1/4$-BPS count finds an interpretation of a motivic stable pair count on an abelian surface, paralleling \cite{KKP}.  Here, we focus on giving the interpretation for the more informative refined $1/8$-BPS count. 

\medskip
Hence, consider some abelian threefold $X$ that splits as a product $A \times E$, with $A$ an abelian surface and $E$ an elliptic curve. In general, for any curve class $\beta \in H_2(X; \mb{Z})$ and some integer $n$ corresponding to our $D0$ number, we may hope to define a Donaldson-Thomas invariant as a (weighted) Euler characteristic of some Hilbert scheme $\Hilb^n(X, \beta) = \{ Z \subset X | [Z] = \beta, \chi(\mc{O}_Z) = n \}$, but as this invariant would essentially always vanish due to the free $X$-action, it is more prudent to quotient by the $X$-action and consider the invariants of the resulting space. We hence conjecture that in our cases of interest, we have some natural quotient $[\Hilb^n(X,\beta)/X] \in K^{\hat{\mu}}(St)[\mb{L}^{-1}]$ in a version of a Grothendieck group of Deligne-Mumford stacks, whose Poincar\'{e} polynomials $P_u$ (as a motivic measure on the Grothendieck group) assemble into a generating function as follows: \be \sum P_u [\Hilb^n(X, (\beta_h, d)) / X] p^d (-y)^n q^h ~=~ \Phi^{\rm refined}_{\rm 4d}~.\ee 

Note that, as in \cite{KT}, we have left the choice of orientation needed to define motivic Donaldson-Thomas invariants somewhat murky above.   As the orientation is essentially a choice of spin structure on the relevant moduli spaces of sheaves, we believe that all relevant moduli spaces in this case have trivial dualizing complex, in the appropriate sense, and that there is consequently a preferred ``zero'' orientation, which moreover happens to be the physically relevant one.  A good physical understanding of the orientation issue remains to be well understood, to our knowledge.

\medskip
The reduced Donaldson-Thomas invariants of $T^4 \times T^2$ were further discussed recently in 
\cite{Ober}.  These authors in particular conjecture a formula (immediately following their Conjecture 2 on page 10) relating the exponential of the generating function of reduced Donaldson-Thomas invariants to the multiplicative lift of $-\phi_{-2,1}$.
In our notation, their formula is\footnote{In this formula and in all subsequent ones in this section, when $n=m=0$, we take $\ell>0$ as usual.}
\be
{\rm exp}\left( \frac12 \(u \frac{\partial}{\partial u}\)^2 \Phi_{\rm 4d}^{\rm refined} \vert_{u=1}\right)
= \prod_{n\geq 0, m\geq0, \ell} {1\over (1-p^n q^m y^\ell)^{c(nm,\ell)}}~.
\ee
One can easily check that this is consistent with the specialization of our refined results for
relatively prime charges (where we know our formulae to hold).
Taking the logarithm of both sides, we obtain from the right hand side
\begin{equation}
 -\sum_{n \geq 0, m\geq0, \ell} c(nm,\ell) p^n q^m y^\ell {\rm log}(1 - p^n q^m y^\ell)\\
= \sum_{n \geq 0, m\geq0, k\geq1,\ell} c(nm,\ell) {1\over k} p^{nk} q^{mk} y^{\ell k}~,
\label{eq:horse}
\end{equation}
while our formula for the left hand side is % $\Phi_{4d}^{\rm refined}$ is
\be
\Phi_{\rm 4d} = \sum_{n \geq0, m\geq0, \ell} {c(nm,\ell)p^n q^m y^\ell \over (1-p^n q^m y^\ell)^2}~.
\ee
Taylor expanding the denominator we obtain
\be
\Phi_{\rm 4d} = \sum_{n\geq0, m\geq0, k\geq1, \ell} c(nm,\ell) k p^{nk} q^{mk} y^{\ell k}~,
\ee
which looks distinct from the result in (\ref{eq:horse}) until one recalls that we are only matching
the coefficients for relatively prime charges.  This fixes $k=1$, and then the two expressions
coincide as expected.  As noted above, extending to non-coprime charges is sure to be interesting both mathematically, for the correct multiple-cover formula, and physically, as we expect the moduli space to turn non-compact.

\bigskip
\centerline{\bf{Acknowledgements}}
\medskip
We thank G. Oberdieck, N. Paquette, X. Yin, and M. Zimet for helpful conversations, and we thank A. Iqbal, C. Kozcaz, and C. Vafa for explanations of relations to other work.  We also thank N. Paquette for providing helpful commentary on early drafts of this manuscript. The research of S.K. was supported in part by the NSF under grant PHY-1316699. N.B. is supported by an NSF Graduate Fellowship and a Stanford Graduate Fellowship.

\end{document}